\documentclass[review, anonymous]{elsarticle}

\usepackage{lineno,hyperref}
\usepackage{tabularx}
\usepackage{longtable}

\modulolinenumbers[5]

\journal{Internet of Things}

\begin{document}

\begin{frontmatter}

\title{Security Aspects of Internet of Things aided Smart Grids: a Bibliometric Survey}

\author[mymainaddress,mysecondaryaddress]{Jacob Sakhnini}
\ead{jsakhnin@uoguelph.ca}

\author[mymainaddress]{Hadis Karimipour}
\ead{hkarimi@uoguelph.ca}

\author[mysecondaryaddress]{Ali Dehghantanha}
\ead{ali@cybersciencelab.org}

\author[mythirdaddress]{Reza M. Parizi}
\ead{rparizi1@kennesaw.edu}

\author[myfourthaddress]{Gautam Srivastava}
\ead{srivastavag@brandonu.ca}

\address[mymainaddress]{School of Engineering, University of Guelph, Guelph, ON, Canada}
\address[mysecondaryaddress]{Cyber Science Lab, School of Computer Science, University of Guelph, Guelph, ON, Canada}
\address[mythirdaddress]{College of Computing and Software Engineering, Kennesaw State University, GA, USA}
\address[myfourthaddress]{Department of Mathematics and Computer Science, Brandon University, MB, Canada}

\begin{abstract}
The integration of sensors and communication technology in power systems, known as the smart grid, is an emerging topic in science and technology. One of the critical issues in the smart grid is its increased vulnerability to cyber threats. As such, various types of threats and defense mechanisms are proposed in literature. This paper offers a bibliometric survey of research papers focused on the security aspects of Internet of Things (IoT) aided smart grids. To the best of the authors' knowledge, this is the very first bibliometric survey paper in this specific field. A bibliometric analysis of all journal articles is performed and the findings are sorted by dates, authorship, and key concepts. Furthermore, this paper also summarizes the types of cyber threats facing the smart grid, the various security mechanisms proposed in literature, as well as the research gaps in the field of smart grid security. 
\end{abstract}


\end{frontmatter}

\section{Introduction}
\label{sec:Intro}
The smart grid is becoming the power systems of the next generation \cite{smartGridSurvey}. The development of today’s power systems is aimed towards integrating smart meters, sensors, and advanced computing technologies \cite{smartGridProjectsEurope}. This smart grid technology greatly enhances the power generation efficiency and prompts the incorporation of various sources of energy generation into one system \cite{smartGridWindEnergy}. The association of smart meters and sensors along power grid networks allows the generation centers access to real-time power demand information. This information can be used to implement an efficient generation and distribution plan \cite{yang_practical_2018,rouzbahani_smart_2019}. As such, integration of these technologies into power system infrastructure has greatly increased the energy efficiency as well as reduced the price of electricity. 

Several countries are investing in smart grid technologies due to its potential for great economic and social benefits \cite{bansal_smart_2016}. However, utilizing communication networks induces security risks and vulnerability to cyber attacks. Therefore, cybersecurity and detection of cyber attacks is an essential part of smart grid movement. The National Institute of Standards and Technology (NIST), the Energy Expert Cyber Security Platform (EESCP), and the European Commission's Smart Grids Task Force have all highlighted the importance of cybersecurity in the emerging smart grid technologies \cite{the_smart_grid_interoperability_panelsmart_grid_cybersecurity_committee_guidelines_2014,EECSP_recommendations,SGTF_recommendations}. As a result, many studies have been published that propose cybersecurity methods and cyber attack identification. 

Smart grid systems consist of various resources and technologies \cite{ericsson_cyber_2010}. Smart meters are incorporated to collect consumption data for more efficient power distribution. Additionally, interconnection of supervisory control and data acquisition (SCADA) allows for more expanded centralized distribution along large geographical areas \cite{ghalavand_microgrid_2018}. The smart grid also allows for interaction among transmission and distribution grids, building controllers, as well as various sources of energy generation \cite{HK1,karimipour_accelerated_2013}. However, the integration of digital and information technology into the smart grid and the increased complexity of the system increases the possibility of cyber attacks and failures propagating from one system to another \cite{khurana_smart-grid_2010}. As such, there are many challenges accompanying cybersecurity in the smart grid. Some examples include the difficulty modelling the non-linearities and stochasticity of the system, as well as modeling the various types of cyber attacks that can potentially inflict the system.

Many Advanced Persistent Threat (APT) actors and hacking teams are targeting critical infrastructure and services \cite{ad1} ranging from healthcare \cite{ad2} and safety critical systems \cite{ad3} to the smart grid. Furthermore, the rise of Internet of Things (IoT) technology which can be defined as a network of physical devices connected to the internet. The use of such devices can help the smart grid by supporting various network functions in power generation and storage as well as provide connectivity between supplier and consumers \cite{6724103}. The integration of IoT devices in the smart grid also poses additional vulnerabilities to cyber threats \cite{8701687}.

Various methods for cyber attack detection have been proposed in literature. Model based solutions, such as variants of state estimation techniques and statistical-based models, have been suggested \cite{tajer_distributed_2011,shuguang_cui_coordinated_2012}. Furthermore, the use of Kalman filters for measurement estimation has been proposed to detect cyber attacks \cite{rawat_detection_2015,kurt_distributed_2018}. Alternatively, intelligent systems have also been proposed in literature. The use of supervised learning was proposed for detection of false data injection (FDI) attacks \cite{ozay_machine_2016,sakhnini_smart_2019}. While supervised machine learning techniques offer higher accuracy \cite{esmalifalak_detecting_2017}, semi-supervised machine learning techniques may rely on lesser studied features such as spatial and temporal correlation of smart meter measurements \cite{chen_detection_2015}.

Other machine learning based solutions have been proposed including reinforcement learning and deep learning algorithms. The use of Artificial Immune Systems (AIS) coupled with an SVM to detect malicious data was proposed in \cite{zhang_distributed_2011}. Alternatively, the use of deep learning to extract the nonlinear features in electric load data to improve on an interval state estimation-based defense system is also proposed in \cite{wang_deep_2018}. Deep learning is also implemented in real-time detection of false data injection attacks in \cite{he_real-time_2017}. Additionally, deep learning is used to analyze real time measurements from PMUs for cyber attack mitigation in \cite{wei_deep_2016}. Recurrent Neural Networks (RNNs) are also proposed for detection of cyber attacks by observing temporal variations in successive historical data in \cite{ayad_detection_2018}. Furthermore, unsupervised deep learning is also used to propose a scalable intelligent attack detection solution in \cite{karimipour_deep_2019}.

Many Advanced Persistent Threat (APT) actors and hacking teams are targeting critical infrastructure and services \cite{ad1} ranging from healthcare \cite{ad2}and safety critical systems \cite{ad3} to the smart grid. 

The variety in and complexity of cyber threats in the smart grid provoked a large amount of solutions.
Therefore, a bibliometric analysis and summary of the up to date solutions to smart grid cybersecurity is important. Such analysis is also lacking in literature. Several summaries and literature reviews have already been published on the topic. For example, literature review and a survey on smart grid cybersecurity was given in \cite{Baumeister2010LiteratureRO} and\cite{wang_cyber_2013}. Similarly, a systematic mapping study of cyber-physical systems has also been published in \cite{zacchia_lun_cyber-physical_2016}. However, these reviews have all been published before 2016 and, as such, are outdated and do not include many of the new solutions proposed. More recent literature reviews have been published which analyze the various types of cyber threats in the smart grid through a survey of literature \cite{otuoze_smart_2018,mrabet_cyber-security_2018}. Both articles, however, lack a bibliometric analysis of literature as well as an inquiry of the attack detection methods used. There are also surveys of smart grid cybersecurity articles \cite{leszczyna_cybersecurity_2018,leszczyna_review_2018}; both of which have emphasized cybersecurity standards and lack details regarding types of cyber attacks and defense mechanisms.

The purpose of this paper is to identify, classify, and review existing publications on cybersecurity of the smart grid to better understand current security mechanisms in literature. A bibliometric analysis is performed on related articles to date, to categorize the publications by its bibliometric characteristics such as authors and dates. This bibliometric analysis can allow researchers to better understand the state of the art of the cybersecurity systems implemented in the smart grid as well as the structure and characteristics of studies in this field. Understanding patterns in research activities can improve future work and research in the field of smart grid cybersecurity. To perform a successful bibliometric evaluation, this paper aims to investigate journal articles published between January 2010 and May 2019 in the domain of cybersecurity in the smart grid. The paper will consider the following research questions:

\begin{enumerate}
\item[a)] What is the trend of publications in smart grid cybersecurity?
\item[b)] What is the future direction of cybersecurity in the smart grid?
\item[c)] What are the types of cyber threats currently studied?
\item[d)] What are the defence mechanisms used in IoT integrated the smart grid?
\end{enumerate}

The remainder of this paper is organized as follows: Section \ref{sec:Meth} outlines the research methods used for the bibliometric analysis. Section \ref{sec:Findings} demonstrates the resultant findings. Next, Section \ref{sec:ReportedAttacks} discusses the attacks on power systems that have been reported. Section \ref{sec:Discussion} provides classification of cyber attack detection in the smart grid. Finally, Section \ref{sec:Conclusion} is the conclusion of this study and describes the challenges and future trends of smart grid cybersecurity.

\section{Methodology}
\label{sec:Meth}
The methods used in this paper follow a similar process to \cite{razak_rise_2016}, which divides the bibliometric methods into two parts. First, general instructions, which discusses the search methods and search engines used to find papers and eliminate possible sources of error. Next, publication analysis, which discusses the evaluation of the publications. This method of bibliometric analysis is used in various subjects, such as the rise of malware in the cybersecurity domain \cite{razak_rise_2016}, the expansion of scientific literature on women in science and higher education \cite{dehdarirad_research_2015}, and literature trends in the domain of biomass energy \cite{mao_past_2015}. Since this bibliometric method is widely used in various subjects of literature, this paper will follow the same methodology and apply it to the domain of cybersecurity in the smart grid.

Online research databases are used to retrieve all the relevant journal articles from January 2010 to May 2019.  There are various online databases that include papers in this domain. IEEE Explore and ScienceDirect are two of the largest databases for smart grid related publications \cite{zacchia_lun_cyber-physical_2016}. The Web of Science (WoS) database is also commonly used since it includes publications from various international databases \cite{razak_rise_2016}. The three aforementioned databases are used with the following search query, which was chosen based on survey papers in this field \cite{otuoze_smart_2018,mrabet_cyber-security_2018,leszczyna_cybersecurity_2018,leszczyna_review_2018}:

("Smart Grid" AND "Cyber Security" OR "Cyber Attack" OR "Cyber Threat" OR "False Data Injection" OR "Attack Detection")

Results from all databases are cross-referenced for repeated results. The results are filtered based on their relevance to cybersecurity of the smart grid, which is evaluated based on the abstracts of the papers. The papers excluded from the bibliometric analysis include papers written in any language other than English, and papers that contain the specified keywords but are not relevant to smart grid cybersecurity. The included papers are categorized according to timeline, journals, and authors. The papers are also categorized based on the research output, which will be mainly measured by the frequency of key words and phrases. Software such as Zotero and VOSviewer tool are used to sort and visualize the bibliometric data.

\section{Findings}
\label{sec:Findings}
\begin{figure}[htbp]
\centering
\includegraphics[width=3.5in]{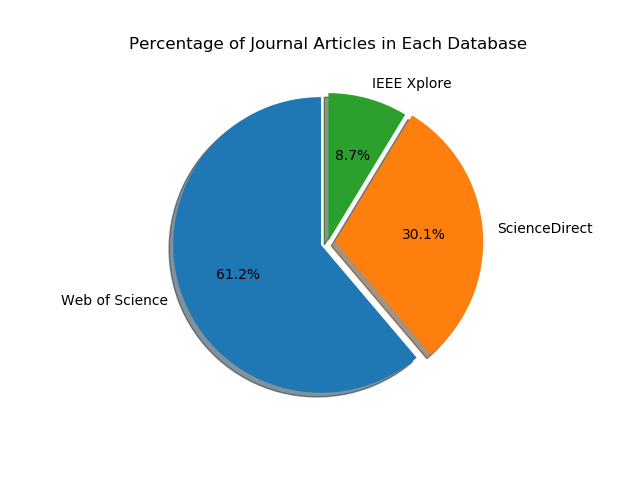}
\caption{Percentage of journal articles published in each database on the topic of security systems in the smart grid}
\label{fig:Databases}
\end{figure}

This section discusses the findings of the bibliometric analysis on the topic of security systems in the smart grid. Using the search query specified in the Section \ref{sec:Meth}, the three databases found a total of $2314$ search results for journal articles. After filtering duplicates, $1722$ journal articles remained. Figure \ref{fig:Databases} shows that the largest number of journal articles on this topic are in the WoS database which account for $61.2\%$ of the findings. ScienceDirect and IEEE Xplore databases have fewer results accounting for $30.1\%$ and $8.67\%$ respectively. Furthermore, duplicates among the three databases accounted for $25.6\%$ which indicates that while some papers are mentioned in multiple databases, the use of only one database is not a sufficient tool to represent the state-of-the-art in this topic.

\begin{figure}[!htbp]
\centering
\includegraphics[width=5in]{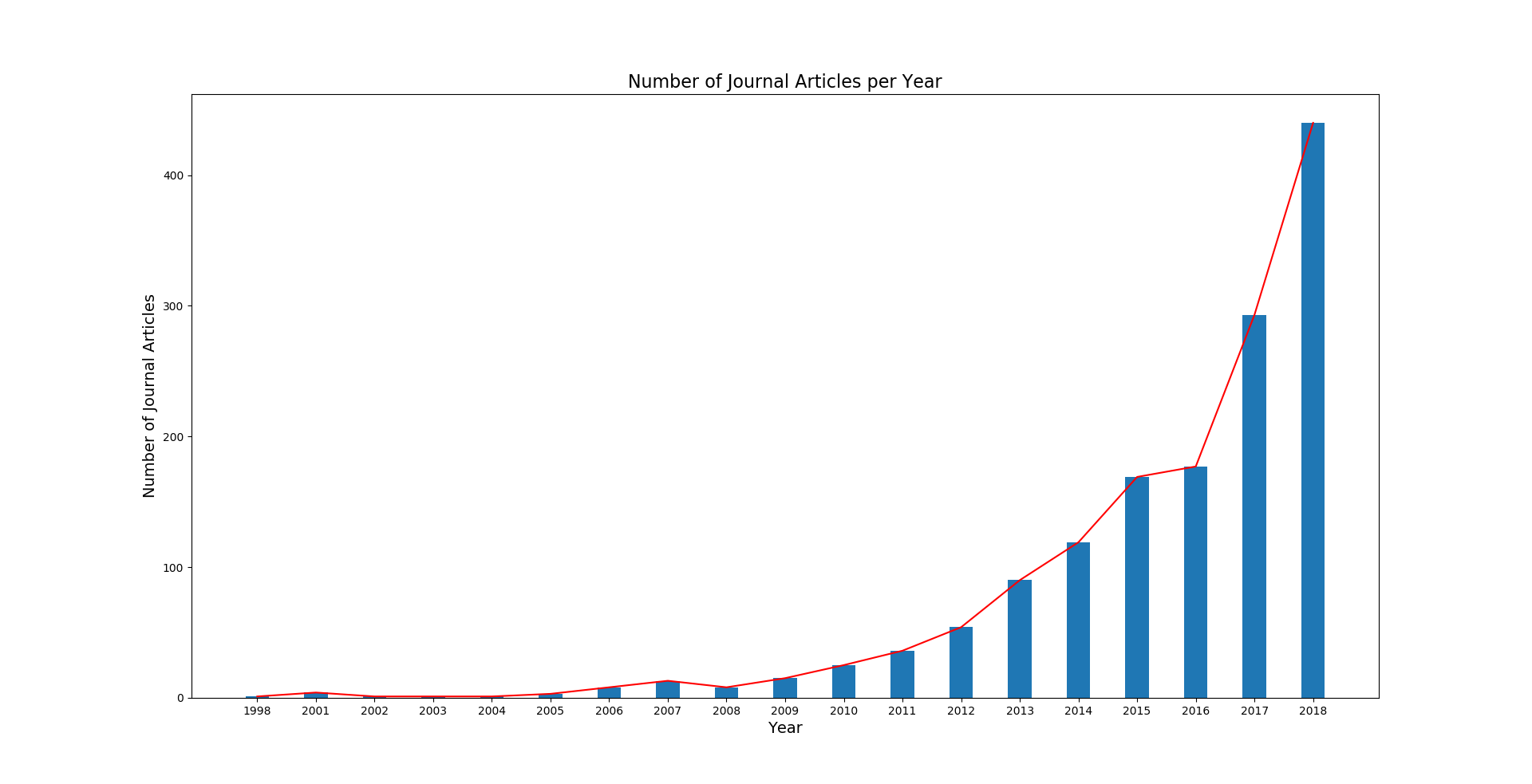}
\caption{Number of journal articles on security systems of the smart grid published every year}
\label{fig:ArticlesPerYear}
\end{figure}

After filtering duplicates, the $1722$ journal articles are sorted by year of publication as seen in Figure \ref{fig:ArticlesPerYear}. The figure demonstrates the novelty of the subject of security systems in the smart grid. Moreover, the majority of the articles were published in the last $5$ years; $30.2\%$ of the journal articles were published in 2018, $20.1\%$ were published in 2017, $12.1\%$ in 2016, and $8.2\%$ and $11.6\%$ were published in 2014 and 2015 respectively. Furthermore, the upward trend in publications over time demonstrates substantial growth in this research topic. Additionally, the small number of publications prior to 2010 show that security systems in the smart grid is a field of study that only recently commenced; meaning that there are likely several uncertainties and much to explore in this research field. 

To study the contributions of different journals to the topic of security systems in the smart grid, the articles are sorted by journal of publication. Table \ref{tab:ArticlesPerJournal} shows the $40$ most prominent journals in this research field, sorted by number of publications. These results demonstrate that while some journals hold more publications in this field than others, the research papers are spread out among a large number of journals. The wide distribution of articles among journals proves that this research area is of general importance to science and society and many journals seek publications in this topic.

\begin{longtable}{c|c|c}
	\caption{Number of articles published in each journal}\\
    \hline
    \textbf{Journal Name} & \textbf{Articles}&\textbf{Percentage}\\
    \hline
    Future Generation Computer Systems&68&$3.95\%$\\
    Computers \& Security&66&$3.83\%$\\
    IEEE Access&52&$3.02\%$\\
    IEEE Transactions on Smart Grid&43&$2.50\%$\\
    IFAC-PapersOnLine&40&$2.32\%$\\
    The Electricity Journal&36&$2.09\%$\\
    Security and Communication Networks&35&$2.03\%$\\
    Computer Networks&34&$1.97\%$\\
    International Journal of Electrical Power \& Energy Systems&34&$1.97\%$\\
    Procedia Computer Science&30&$1.74\%$\\
    International Journal of Critical Infrastructure Protection&29&$1.68\%$\\
    IEEE Transactions on Information Forensics and Security&25&$1.45\%$\\
    Ad Hoc Networks&21&$1.22\%$\\
    Energy Procedia&20&$1.16\%$\\
    Sensors&20&$1.16\%$\\
    Journal of Network and Computer Applications&19&$1.10\%$\\
    Neurocomputing&19&$1.10\%$\\
    Electric Power Systems Research&18&$1.05\%$\\
    IFAC Proceedings Volumes&17&$0.99\%$\\
    Computer Communications&16&$0.93\%$\\
    Computers \& Electrical Engineering&16&$0.93\%$\\
    Energy Policy&15&$0.87\%$\\
    Expert Systems with Applications&15&$0.87\%$\\
    IEICE Transactions on Information and Systems&15&$0.87\%$\\
    Wireless Personal Communications&15&$0.87\%$\\
    Journal of the Franklin Institute&14&$0.81\%$\\
    Information Sciences&13&$0.75\%$\\
    International Journal of Distributed Sensor Networks&13&$0.75\%$\\
    Procedia Technology&13&$0.75\%$\\
    Automatica&12&$0.70\%$\\
    IEEE Communications Surveys \& Tutorials&11&$0.64\%$\\
    IEEE Transactions on Industrial Informatics&11&$0.64\%$\\
    Multimedia Tools and Applications&11&$0.64\%$\\
    IEEE Transactions on Power Systems&10&$0.58\%$\\
    Journal of Information Security and Applications&10&$0.58\%$\\
    Sustainable Energy, Grids and Networks&10&$0.58\%$\\
    Energy&9&$0.52\%$\\
    International Journal of Security and Its Applications&9&$0.52\%$\\
    Technological Forecasting and Social Change&9&$0.52\%$\\
    Energies&8&$0.46\%$
    \label{tab:ArticlesPerJournal}
\end{longtable}

Aside from journals, articles are also sorted and analyzed based on author. Table \ref{tab:authorTable} shows the $40$ authors with the most publications in security systems of the smart grid. The author with the most publications in this research field, Kim-Kwang Raymond Choo, has authored $1.16\%$ of the publications found. The authors with the second and third most publications, Xuan Liu and Zuyi Li, have authored $0.58\%$ and $0.52\%$ of the journal articles respectively. The low percentage of publications for the most prominent authors in the field shows that security systems in the smart grid is a vastly growing field that is gaining interest from various authors around the world. Furthermore, there are $4952$ authors who contributed to the $1722$ journal articles analyzed. This demonstrates the high demand for research advancement in the topic of security systems in the smart grid.

\begin{longtable}{c|c|c}
    \caption{Number of articles published by each author}\\
        \hline
        \textbf{Author}&\textbf{Publications}&\textbf{Percentage}\\
        \hline
        Choo, Kim-Kwang Raymond&20&$1.16\%$\\
        Liu, Xuan&10&$0.58\%$\\
        Li, Zuyi&9&$0.52\%$\\
        Abbas, Haider&8&$0.46\%$\\
        Shon, Taeshik&8&$0.46\%$\\
        Genge, Béla&7&$0.41\%$\\
        Kumar, Neeraj&7&$0.41\%$\\
        Lopez, Javier&7&$0.41\%$\\
        Lu, Rongxing&7&$0.41\%$\\
        Xiao, Yang&7&$0.41\%$\\
        Xu, Shouhuai&7&$0.41\%$\\
        Das, Ashok Kumar&6&$0.35\%$\\
        Debbabi, Mourad&6&$0.35\%$\\
        Govindarasu, Manimaran&6&$0.35\%$\\
        Haller, Piroska&6&$0.35\%$\\
        Shi, Ling&6&$0.35\%$\\
        Alcaraz, Cristina&5&$0.29\%$\\
        Elovici, Yuval&5&$0.29\%$\\
        Guizani, Mohsen&5&$0.29\%$\\
        Han, Zhu&5&$0.29\%$\\
        Janicke, Helge&5&$0.29\%$\\
        Kundur, Deepa&5&$0.29\%$\\
        Liu, Anfeng&5&$0.29\%$\\
        Qian, Yi&5&$0.29\%$\\
        Qiu, Meikang&5&$0.29\%$\\
        Vasilakos, Athanasios V.&5&$0.29\%$\\
        Wang, Jianhui&5&$0.29\%$\\
        Wang, Wei&5&$0.29\%$\\
        Wazid, Mohammad&5&$0.29\%$\\
        Xiang, Yang&5&$0.29\%$\\
        Yu, Wei&5&$0.29\%$\\
        Amjad, Muhammad Faisal&4&$0.23\%$\\
        Anwar, Zahid&4&$0.23\%$\\
        Bou-Harb, Elias&4&$0.23\%$\\
        Bretas, Arturo S.&4&$0.23\%$\\
        Cai, Zhongmin&4&$0.23\%$\\
        Chen, Jiming&4&$0.23\%$\\
        Chilamkurti, Naveen&4&$0.23\%$\\
        Czajka, Adam&4&$0.23\%$\\
        Du, Xiaojiang&4&$0.23\%$
        \label{tab:authorTable}

\end{longtable}

\begin{figure*}[!htbp]
    \centering
    \includegraphics[width=5in]{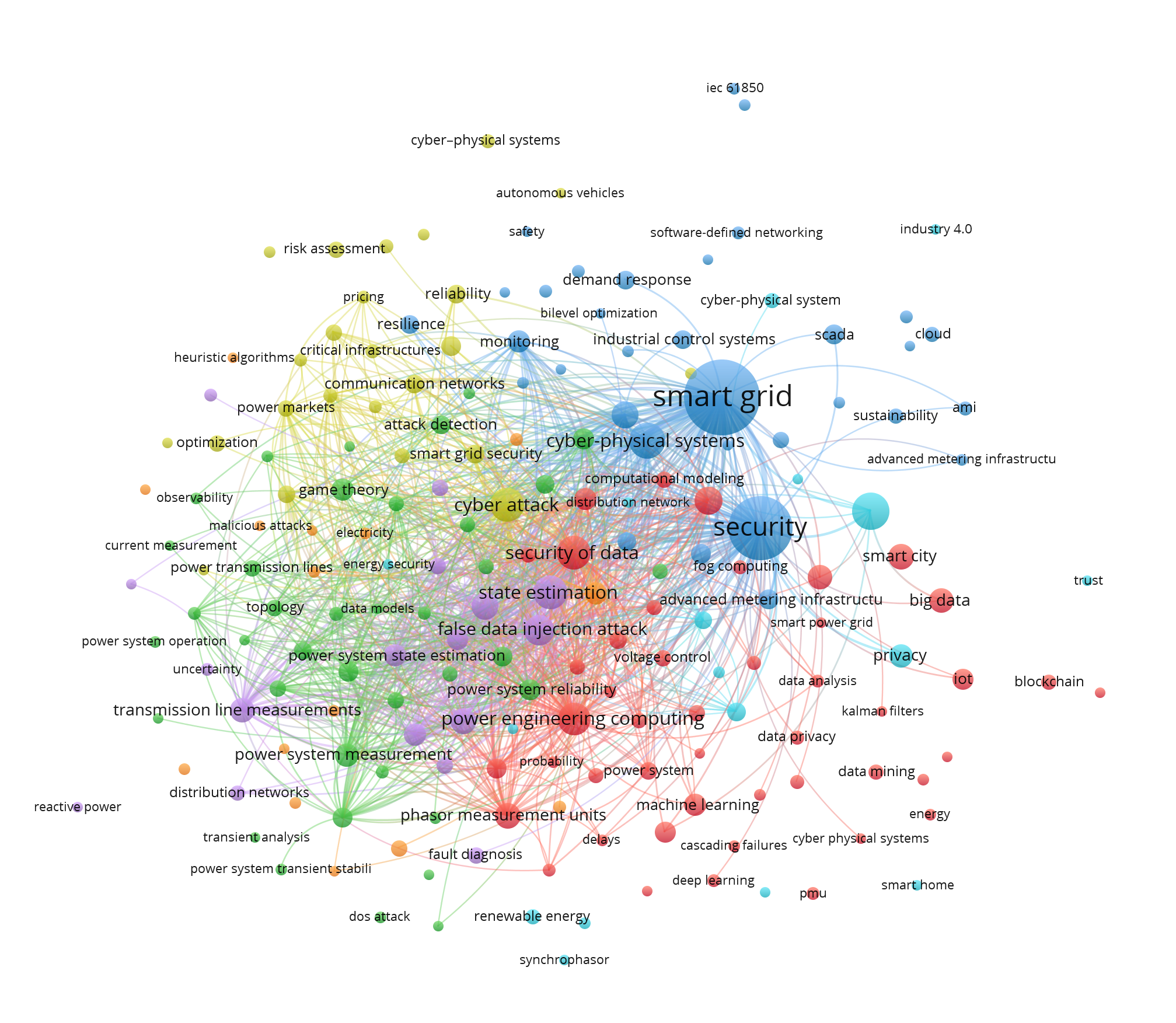}
    \caption{A heat-map of keywords mentioned in all journal articles in the subject of smart grid cybersecurity}
    \label{fig:heatmap}
\end{figure*}

To explore the specific aspects of smart grid cybersecurity, a keyword analysis was performed. The heat-map in 
Figure \ref{fig:heatmap} shows the most prominent keywords observed in the articles analyzed. This heat-map represents the keywords in colored clusters as well as their connections. Figure \ref{fig:heatmap} exhibits the most significant topics in this subject based on their occurrence in journal papers. This heatmap shows the most important terms surrounding smart grid security which demonstrates the main concepts associated with this research field.

The findings of this bibliometric analysis demonstrate the significance of security in the smart grid. The number of contributing authors and journals show an active community of researchers focused on this topic. Additionally, the increasing number of publications in recent years shows a great increase in the interest of this topic. To further demonstrate the significance of security in the smart grid, the next section discusses reported attacks on power systems in recent history as well as their impact.

\section{Reported Attacks on the Smart Grid}
\label{sec:ReportedAttacks}
There have been several documented attacks on the electric grid attributed to cyber attacks. In January 2003, the computer network at the Davis-Besse nuclear plant in Oak Harbor, Ohio was compromised by a malware disabling its processing computer and safety monitoring system for several hours \cite{office_cybersecurity:_2012}. Similarly, circulation pumps at the Brown Ferry nuclear plant in Alabama failed due to excessive traffic, believed to be attributed to a DoS attack \cite{office_cybersecurity:_2012}. Furthermore, an investigation in 2009 revealed that hackers are able to steal power through compromising the smart meters and changing the consumption readings \cite{goel_chapter_2015}. Phishing incidents have also been reported at electric bulk providers and malware samples were found indicating a targeted and sophisticated intrusion \cite{office_cybersecurity:_2012}.  Additionally, in April of 2012, the FBI was asked to investigate widespread incidents of power thefts through smart meter attacks \cite{goel_chapter_2015}. The report indicates that hackers changed the power consumption of smart meters using software available easily on the internet. 

Such incidents in recent history induce various security concerns regarding critical infrastructure. As such, it is crucial that security of the smart grid is explored at every level including adequate situational awareness at all times. In fact, lack of situational awareness can have devastating impacts beyond cyber threats. For example, in August of 2003, a blackout occurred in the north east of the United States due to a cascading failure of the power system due to the lack of awareness of the Ohio-based electric utility company. This lack of awareness resulted in a cascading failure of $508$ generators and $265$ power plants across eight states and southern Ontario \cite{goel_chapter_2015}. This clearly shows how adequate security systems can have benefits beyond mitigating cyber threats, including minimizing damage from faults or incidents.

\section{Security Systems in the Smart Grid}
\label{sec:Discussion}
This section examines the security threats facing the smart grid, as well as the state of the art of the current countermeasures of these threats. Subsection \ref{ssec:Disc_Threats} discusses the specific types of cyber threats in power systems, and subsection \ref{ssec:Disc_SecurityMeasures} discusses the defense mechanisms proposed in literature.

\subsection{Types of Cyber Threats}
\label{ssec:Disc_Threats}
Cyber threats or cyber attacks are among the most discussed and studied threats for the smart grid \cite{otuoze_smart_2018}. The wide interest in studying cyber threats in the smart grid is due to the number of significant vulnerabilities identified \cite{mcdaniel_security_2009}. Furthermore, cyber attacks have the potential of leading power systems into total collapse \cite{goel_security_2015}. These cyber attacks can occur for various purposes and are generally divided into two main types: Passive Attacks and Active Attacks \cite{delgado-gomes_smart_2015}. Passive attacks include eavesdropping, spying, and traffic analysis; while active attacks include denial of service (DoS) and malware attacks.

\begin{figure}
    \centering
    \includegraphics[width=6in]{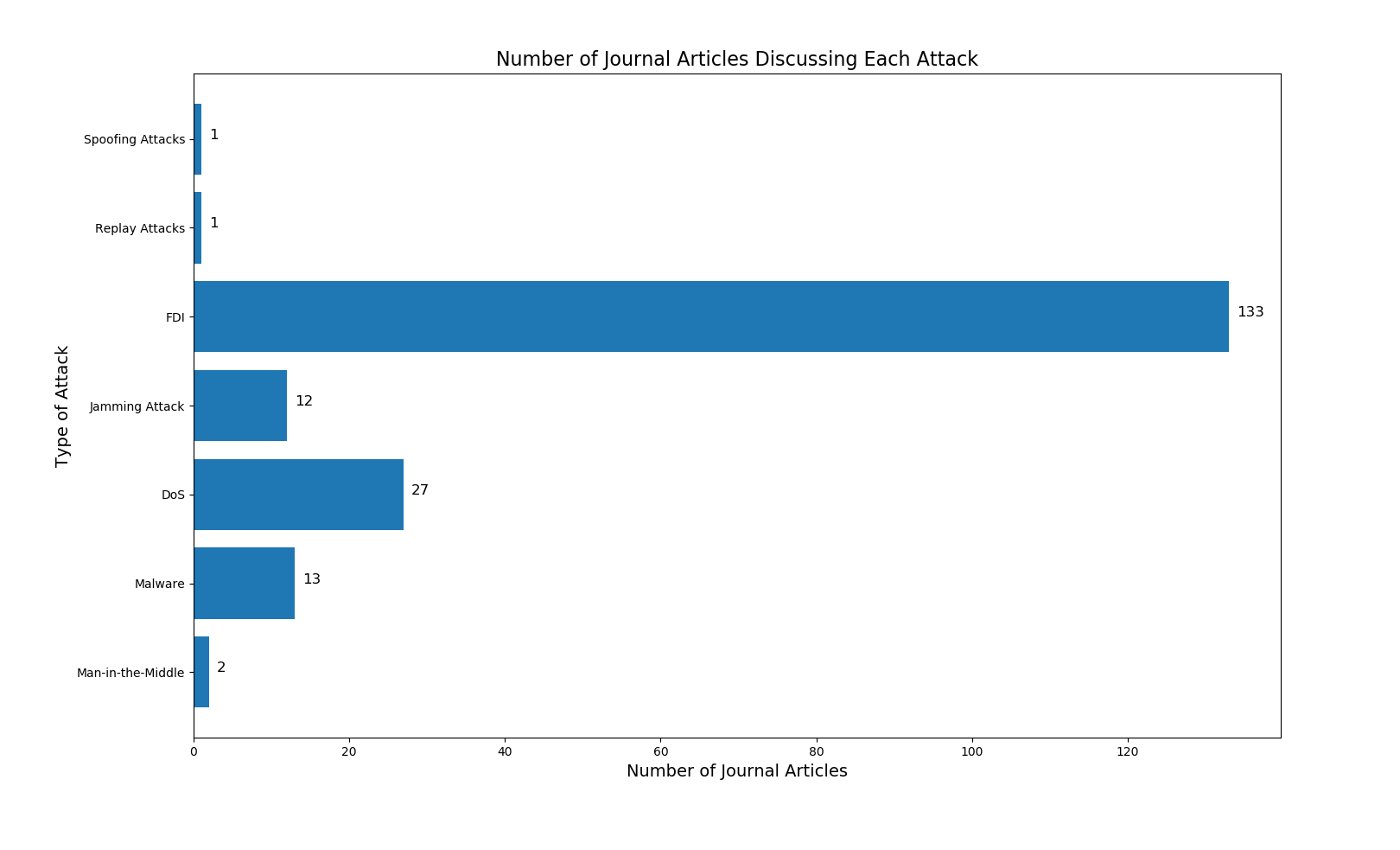}
    \caption{The number of journal articles studying each attack type}
    \label{fig:attackTypes}
\end{figure}{}

The various types of attacks are not equally studied in literature. Figure \ref{fig:attackTypes} shows the number of articles studying each type of attack. While there are more types of cyber threats that can compromise a network, the following sections discuss the attacks studied in the smart grid, which are mentioned in Figure \ref{fig:attackTypes}.

\subsubsection{Spoofing Attacks}
\label{sssec:Spoofing}
The main types of spoofing are GPS spoofing, ARP spoofing, and IP spoofing \cite{jokar_spoofing_2013}. IP spoofing uses a modified IP to pass through security systems and is typically the first stage of a complex intrusion. GPS spoofing, however, is based on broadcasting incorrect signals of higher strength than received from satellites to deceive victims. ARP spoofing is where falsified ARP (Address Resolution Protocol) messages are used to link the attacker’s MAC address with the IP address of the victim. Through this all data in the compromised system will pass through the intruder. The most common type of spoofing attacks in the smart grid is GPS spoofing due to the use of GPS receivers in the metering infrastructure. Vulnerability analysis in literature demonstrates how phasor measurement units (PMUs) are susceptible to GPS spoofing attacks \cite{risbud_vulnerability_2018}. GPS spoofing attacks can mislead the network operator, and drastically impact subsequent corrective control actions \cite{7860525}. 

\subsubsection{Replay Attacks}
\label{sssec:ReplayAttacks}
Replay attacks aim to intercept authentication information. In the smart grid, replay attacks intercept the usage pattern along the varying smart meters and replay this data to carry out an undetected intrusion \cite{7820542}. The integration of IoT devices in smart grid networks induces increased threat to these attacks. Furthermore, attacker can inject incorrect data to the system, which may lead to incorrect energy price or inaccurate prediction \cite{6482409}.

\subsubsection{Man-in-the-Middle Attack}
\label{sssec:ManInMiddle}
This attack makes use of ARP, which maps a protocol address to a hardware address (MAC address) \cite{jinhua_arp_2013}. The purpose of this attack is to combine the attacker’s MAC address with the host’s IP address triggering any traffic meant for that particular IP to be sent to the attacker instead, this is referred to as ARP spoofing \cite{sharma_detection_2014}. This allows the attacker to capture the communication information within the SCADA system \cite{yang_man---middle_2012}.

\subsubsection{Smart Meter DoS Attacks}
\label{sssec:DOSattacks}
DoS attacks are typically achieved by flooding specific nodes of the system with data that prompts generating and sending large volume of reply and request packets \cite{yi_denial_2014}. There are various methods for generating such attacks which can cause a system blackout \cite{guo_modeling_2015}. These attacks can also be implemented through IoT devices integrated into the smart grid. The increased integration of these IoT devices has led to increased interest in DoS attacks \cite{BEKARA2014532}.

\subsubsection{Malware}
\label{sssec:malware}
The propagation of malicious software, known as malware, is another potential cyber threat to the smart grid \cite{Dovorm_fuzzy}. One paper tests security methods for three different types of malware, pandemic malware, endemic malware, and contagion malware \cite{malware_smartgrid}. Pandemic malware is an aggressive malware that infects all devices in the shortest time possible through a topological scan strategy. Endemic malware, however, is the more intelligent type which sacrifices speed for stealth by operating with less conspicuous hit list. Finally, contagion malware is highly stealth and does not initiate connections with the network but rather appends on legitimate communication flows.

The dangers of malware are accentuated in the communication layer of the smart grid. It has been noted that thousands of smart meters may feature identical hardware and firmware \cite{Eder-Neuhauser2018}. While this reduces cost and automates maintenance, the closeness in device types and software induces susceptibility to malware propagation.

\subsubsection{False Data Injection Attacks}
\label{sssec:FDIattacks}
False Data Injection (FDI) attacks consist of malicious data injected into measurement meters \cite{tian_data-driven_2018}. FDI attacks can be performed by manipulating the measurements along the network by a linear factor of the Jacobian matrix of the power system \cite{sakhnini_smart_2019,karimipour_false_2017}. This change in measurement is undetected by the current state estimation techniques \cite{liu_false_2017-1}. Furthermore, these attacks can be created in various strategies with limited knowledge of power system topology \cite{liu_false_2017,wang_novel_2017,zhong_novel_2018}. As such, these types of attacks are widely studied in the smart grid cybersecurity field \cite{tian_data-driven_2018,liu_false_2017-1,liu_false_2017,wang_novel_2017,zhong_novel_2018,wang_false_2018,lei_false_2016,che_false_2019,kang_false_2018,li_false_2018,chai_impacts_2014,liu_jamming_2014}.

\subsubsection{Micro-Grid-Based Jamming Attack}
\label{sssec:JammingAttack}
This type of attack consists of jamming specific signal channels to intervene and disrupt data transmission \cite{gai_spoofing-jamming_2017}. This results in unreliable communications and decreased performance in the power system \cite{tazi_review_2015,ericsson_cyber_2010}.

\subsection{Defence Mechanisms}
\label{ssec:Disc_SecurityMeasures}
Security and defense against the aforementioned attacks and threats is achieved through various mechanisms. The security measures proposed in literature are divided into the ``7D model" or the 7 phases of cybersecurity as given in  \cite{mwiki_analysis_2019}:
 
\begin{itemize}
\item discovery
\item detection
\item denial
\item disruption
\item degradation
\item destruction
\end{itemize}

The following subsections of Subsection \ref{ssec:Disc_SecurityMeasures} will discuss each of the components and their proposed methods in literature.

\subsubsection{Discovery}
\label{sssec:Discovery}
The discovery process in cybersecurity involves identifying and locating sensitive data for adequate protection. In general applications of cybersecurity, data discovery consists of auditing regulated information to ensure its protection. This is helpful because it enables context aware security, in which information within the system is protected based on its sensitivity. In smart grid security, the discovery phase mainly consists of identification of vulnerabilities within the system.

Various methods are proposed in literature for vulnerability analysis in power grids. One paper proposes an automated binary-based vulnerability discovery method that extracts security-related features from the system \cite{kwon_automated_2017}. This automatic discovery algorithm is tested on real smart meter data from Korean infrastructure. Vulnerability analysis specific to certain types of environments or threats are also proposed. In another paper, the survivability of smart grid under is modeled under random and targeted attacks considering a networking islanding scheme \cite{chopade_modeling_2012}. Another paper uses automatic static analysis (ASA) to detect buffer-overflow vulnerabilities of terminal devices \cite{ying_detecting_2019}. Such automated techniques for vulnerability analysis can be useful due to their robustness and scalability to larger systems. As such, a comprehensive assessment of vulnerabilities in the smart grid from past to future has been published highlighting the various vulnerabilities and discovery techniques \cite{chen_complex_2012}.

More specific vulnerability modeling is also proposed in literature. One such work models the vulnerabilities of the smart grid with incomplete topology information \cite{srivastava_modeling_2013}. The results of this paper demonstrate the high level of threat in the smart grid by exhibiting vulnerabilities that can be exploited with limited knowledge of the system. Another paper reveals the cascading failure vulnerability in the smart grid using a novel metric, called risk graph, which shows the importance of nodes within the system as well as the relationship among them \cite{zhu_revealing_2014}. Using this method, Zhu \textit{et al.} develop a new node attack strategy and introduce new vulnerabilities not considered before in literature. 

Vulnerability analyses are also performed on specific attacks. One paper performs a vulnerability analysis of the smart grid to GPS spoofing, a type of attack capable of altering measurements to mislead network operators \cite{risbud_vulnerability_2018}. Another paper analyzes the vulnerability for simultaneous attacks in the smart grid \cite{paul_vulnerability_2017}. Paul and Ni consider various combinations of attacks and proposes a new damage measurement matrix to quantify the loss of generation power and time to reach steady-state. Web-based threats are also considered in another paper which tackles the penetration of digital devices in the smart grid and the associated consequences \cite{dehalwar_review_2017}.

Most articles assess the vulnerability of the smart grid by analyzing either substations or transmission lines. One article, however, performs a vulnerability assessment on a joint substation and transmission line system in which attacks can happen in either the substation, the transmission line, or both \cite{zhu_joint_2015}. Another article takes into account scenarios of severe emergencies in the smart grid and SCADA network and performs a vulnerability analysis of the system under emergencies such as attacks from weapons of mass destruction (WMD) \cite{chopade_structural_2013}. Chopade and Bikdash analyze structural vulnerabilities, which consider infrastructures topology, and functional vulnerabilities, which consider operating regimes of different infrastructures. 

As demonstrated by the aforementioned articles, there is sufficient analysis on vulnerabilities in the smart grid. Various attack strategies are identified and implemented in literature that demonstrate the potential of cyber threats. Detection and mitigation of some of these threats remains as a gap in this research field. Next, we discuss the detection mechanisms proposed as well as the future trend in these methods.

\subsubsection{Detection of Attacks}
\label{sssec:Detection}
Detection of cyber threats is typically achieved through classification using data or measurements across the power system. Measurements along various infrastructure and communication layers of the system are used to detect the presence of threats or attacks. Model-based techniques are used to detect cyber attacks through meter measurements through enhanced state-estimation techniques \cite{tajer_distributed_2011,karimipour_robust_2018,karimipour_parallel_2015}. Furthermore, distributed algorithms are used to find statistical variations in cyber attack vectors \cite{rawat_detection_2015}. Kalman filters are also used to estimate measurements along the power system along with statistical methods of finding anomalies in measurements \cite{rawat_detection_2015,kurt_distributed_2018,karimipour_extended_2015}. 

For defense methods to be scalable to larger systems, purely model-based attack detection techniques are insufficient to guarantee the security of the smart grid \cite{yilin_mo_cyberphysical_2012,sakhnini_smart_2019}. As such, the use of intelligent systems and machine learning for detecting cyber attacks is proposed. Supervised and unsupervised learning have been tested and compared to conclude that supervised learning approaches generally result in more accurate classification of attacks \cite{SVManomaly}. Various supervised learning algorithms have been successfully implemented \cite{OzayML,SVMandKNN}. The results of comparing these learning algorithms demonstrate that a Gaussian-based Support Vector Machine (SVM) is more robust with more accurate classification among larger test systems \cite{SVMandKNN}. Furthermore, another paper implemented the margin setting algorithm (MSA) demonstrating better results than SVM and artificial neural networks (ANN) \cite{8093999,sakhnini_smart_2019}. Other intelligent techniques include adaboost, random forests, and common path mining method \cite{6032057,6982207,7063234}.

A critical concern in the use of intelligent systems in smart grid is computational efficiency \cite{R,mohammadi_cyber_2019}. Many researchers try to tackle this issue by reducing the dimensions of the data through principal component analysis \cite{SVManomaly,OzayML}. One paper proposes the use of a genetic algorithm to select an ideal subset of features that can increase the computation speed while maintaining the detection accuracy of the machine learning algorithms \cite{8357769}. Exploring various feature selection techniques can be effective at increasing the computational efficiency of machine learning algorithms. However, there have not been many papers exploring this subject in the area of smart grid cybersecurity. As such, deep learning techniques with automated or unsupervised feature selection methods are likely to be proposed to tackle the computational burden of larger power systems.

\subsubsection{Denial of Attacks}
\label{sssec:Denial}
One of the security methods in the smart grid revolves around the denial or prevention of cyber threats. Denial techniques pertaining the security of the smart grid typically take the shape of encryption methods for secure communications within the system \cite{tazi_review_2015,malina2019secure}. The most common encryption methods are the use of symmetric or asymmetric keys. Symmetric keys use the same key to encrypt and decrypt the messages while asymmetric keys use different keys for encryption and decryption \cite{ericsson_cyber_2010,dwivedi2019decentralized}. Asymmetric key encryption requires a larger computational capacity and is therefore not suitable for time-sensitive information. Symmetric key encryption does not induce significant computational delay. However, it requires a public infrastructure for key management. Therefore, it is suitable for encryption of distribution and transmission systems \cite{metke_security_2010,dwivedi2019differential,dwivedi2018differential}.

Various encryption and key management methods have been proposed. One scheme is based on Needham-Shroeder authentication protocol and elliptic curve cryptographic algorithms for generating public keys \cite{wu_fault-tolerant_2011}. Another scheme uses digital certificates to establish symmetric communication sessions \cite{metke_security_2010}. Additionally, another authentication method is proposed that is based on S/key one-time password scheme aimed to provide mutual authentication between the meters and servers of the smart grid \cite{leea_s_key-like_2014}. Mutual authentication between smart grid utility network and Home Area Network (HAN) smart meters is also explored through a novel key management protocol \cite{nicanfar_smart_2011}. The proposed mechanism aims at preventing various attacks including Brute-force, Replay, Man-in-The-Middle, and Denial-of-Service attacks. Furthermore, encryption of specific variables and measurements is also studied, specifically pertaining to FDI attacks \cite{khanna_feasibility_2016,li_pama:_2018,li_pama:_2018}. 

Choosing appropriate key management schemes is done by considering the trade-off between security and computational efficiency. However, other issues pertaining denial of attacks arise from the distributed nature of smart grid systems. One paper proposes an efficient framework to read isolated smart grid devices that satisfies the hardware constraints while maintaining integrity against most typical attacks \cite{sha_secure_2017}. Another protocol is proposed for preserving privacy through aggregation of metering data in distributed scenarios and encryption of measurements using a secret sharing scheme \cite{rottondi_implementation_2012}.

Other denial techniques are proposed in literature include increasing situational awareness to prevent attacks. One paper proposes specific measures to tackle issues that lead to lack of awareness among smart grid operators. Such measures include separate networks for actuators and sensors and restricting the use of real time clocks to write-only data storage \cite{shovgenya_demand_2015}. Another paper proposes a different proactive defense approach which consists of randomizing meter infrastructure configurations to lower the predictability of the system to potential adversaries \cite{ali_randomizing_2013}.  While there are many approaches to deny or prevent cyber threats, further research is likely necessary due to the continuous improvement and modifications of adversarial techniques.

\subsubsection{Disruption of Attacks}
\label{sssec:Disruption}
A critical part of the security of any system is the disruption of cyber threats once the system is infected. Disruption of attacks in the smart grid is typically tackled by game theory approaches. One paper demonstrates disruptive countermeasures to reduce the impact of attacks based on the knowledge of non-compromised components \cite{srikantha_attack-mitigation_2016}. Similarly, another paper demonstrates how informed decisions can be made in real-world scenario of attacks to mitigate or disrupt them \cite{hewett_smart_2014}. This is done by using a sequential two-player game model that includes attacker/defender behavior. Similarly, another article attempts to achieve the same goal by making use of the Stackelberg competition, which quantitatively analyzes the game process between attacker and operator \cite{ni_design_2017}. A linear game framework is also proposed with the emphasis on application to large power systems with large number of components under attack \cite{ranjbar_linear_2019}.

Disruption of attacks through game theory is also studied under varying circumstances. One article considers coalitional attacks that can be launched by multiple adversaries \cite{yang_game-theoretic_2016}. A game-theoretic model is proposed to capture the interaction among the adversaries and quantify the capacity of the defender based on Iterated Public Goods Game (IPGG) model. Similarly, stochastic games for protection against coordinated attacks is also proposed in \cite{wei_stochastic_2018}. This method uses an optimal load shedding technique to quantify physical impacts of coordinated attacks which are used as input parameters to model interactions between attacker and defender. Another paper looks into specific types of attacks that exploit cyber vulnerabilities of specific meters and spread into the physical components of the system \cite{shange_game-theory_2014}. This paper also proposes game theory to analyze such attacks. Similarly, a game-theoretic perspective of data injection attacks with multiple adversaries is also studied \cite{sanjab_data_2016}.

There is also focus on the disruption of specific common attacks in the smart grid. Game theory based defense strategies against DoS attacks are proposed which use Nash Equilibrium to maintain dynamic stability in an attacked system \cite{rani_game_2013,srikantha_denial_2015}. Minimizing the effects of jamming attacks is also studied through a modified version of contract network protocol (CNP) as a negotiation protocol among agents \cite{ma_multiact_2015}. Results of this paper indicate that applying the proposed protocol can reduce the jammer’s illegal profit and decrease their motive. The problem with most of the proposed game theory techniques, however, is their tendency to view network interdictions as one-time events. Further research in this topic is likely to take shape as more comprehensive modeling of network interdictions occurs. There are few papers in literature that take this into consideration. One paper, however, uses zero-sum Markov games and a more comprehensive model of attacker behavior \cite{ma_markov_2013}. This paper also demonstrates a defender can use deception as a defense mechanism. Next, we discuss the deception techniques proposed in literature, which when combined with the aforementioned disruption techniques, can act as a comprehensive strategy for mitigating attacks.

\subsubsection{Deception of Attackers}
\label{sssec:Deception}
While disruption of attacks involves minimizing the damage of cyber attacks, deception focuses on altering the direction of the attack to mitigate its impact. This is done by deceiving the attacker into targeting a trap. This deception technology is an emerging field in cyber security due to its potential to detect and defend against zero-day and advanced attacks. In the security of the smart grid, however, deception technology is seldom used in literature.

A strategic honeypot game model was proposed for DoS attacks in the smart grid \cite{wang_strategic_2017}. This paper introduces honeypots into the metering infrastructure network as a decoy system to detect and gather information. Interactions between attackers and defenders are analyzed and the existence of several Bayesian-Nash equilibriums is proved. However, this method was designed and tested for one specific type of attack. A more general honeypot system is proposed to emulate an entire smart grid field communication infrastructure in \cite{mashima_towards_2017}. This paper claims that their honeynet system can emulate high-fidelity and realistic power grid behavior to deceive the attackers. However, evaluation of its realism and scalability are only preliminary and testing was done on a single simulated system. Another paper identifies the various types of honeypots and built a test system to emulate a device on a utility network \cite{hastings_tracking_2014}. However, similarly to the aforementioned papers, analysis regarding realism and scalability are insufficient. This is identified as a research gap in the deception strategies for smart grid security. Future research is expected to involve more comprehensive system modeling and the proposal of more versatile honeynet systems.

\subsubsection{Degradation or Destruction of Attacks}
\label{sssec:DestroyDegrade}
Degrading or destroying the attack is the final part of the defense strategy in the smart grid and it involves minimizing or destroying the effects of the attack. An example of such mitigation techniques include defining security metrics that quantify the importance of individual substations \cite{vukovic_network-aware_2012}. Another proposed method uses a distinctive modeling technique with the capability to modify network topology \cite{delgadillo_analysis_2010}. Such a technique can be used to degrade the attack through optimizing the operation of the power system to minimize its effects. This is done through a mixed-integer nonlinear bi-level program; in the upper-level a terrorist agent maximizes the damage caused in the power system, and in the lower level the system operator minimizes the damage through optimal operation of the power system. Furthermore, the paper proposes a Benders decomposition approach to transform the problem into a standard one-level optimization problem. Another paper, however, tackles the same problem through a genetic algorithm \cite{arroyo_genetic_2009}. Alternatively, another paper proposes a different tri-level model for power network defense with the same goal of minimizing economic cost that the attacks may cause \cite{yao_trilevel_2007}.

Degradation techniques are often coupled with disruption techniques in game theory approaches, as mentioned in Subsection \ref{sssec:Disruption}. As such, defense solutions that only focus on degradation of attacks are limited. Furthermore, due to the legal implications, there are no solutions proposed that focus on destroying the attack through hostile actions towards the adversary. Therefore, most solutions in literature focus on denying, detecting, and minimizing the effect of attacks.

\section{Conclusion}
\label{sec:Conclusion}
This paper analyzes publications within the field of security of IoT-aided smart grids. Bibliometric results are reported highlighting the significance and growth of this field. The findings demonstrate exponential growth in the field of security systems in the smart grid over the last decade. Furthermore, the journal papers analyzed discuss a variety of issues and various types of solutions. The findings are summarized and types of threats, security measures, and evaluation methods of the security systems are discussed. This bibliometric analysis concludes that the variety of threats and the complexity of smart grid systems calls for comprehensive and intelligent security methods. Furthermore, the primary concerns in detection of cyber threats are computational efficiency and minimizing the rate of false positives. Therefore, future publications are expected to take aim at increasing the computational speed of security algorithms while maintaining a high detection accuracy and a low rate of false alarms. Another research gap in this field is the mitigation of cyber threats that have already infected a smart grid. Most papers focus on detection and prevention of cyber threats and only a few papers focus on mitigating those threats. Therefore, future trends in this field of study are projected towards the mitigation of cyber threats as well as robust deep learning algorithms for efficient detection of cyber threats.


\bibliography{mybib}

\end{document}